\documentclass[reprint,amsmath,amssymb, aps,prl]{revtex4-1}

\usepackage[utf8]{inputenc}
\usepackage[english]{babel}
\usepackage{graphicx}
\usepackage{dcolumn}
\usepackage{bm}
\usepackage{braket}
\usepackage{color}

\usepackage{hyperref}
\hypersetup{pdfborder=0 0 0,colorlinks=true,citecolor=blue,linkcolor=blue}

\newcommand{\im}{\operatorname{i}}

\DeclareMathOperator*{\Tr}{Tr}
\allowdisplaybreaks

\begin{document}

\title{Assessing density functionals using many body theory for hybrid 
perovskites }

\author{Menno Bokdam}
\email{menno.bokdam@univie.ac.at}
\author{Jonathan Lahnsteiner}
\author{Benjamin Ramberger}
\author{Tobias Schäfer}
\author{Georg Kresse}
\email{georg.kresse@univie.ac.at}
\affiliation{%
 University of Vienna, Faculty of Physics and Center for Computational Materials
Sciences, Sensengasse 8/12, 1090, Vienna, Austria
}%

\date{Published in Physical Review Letters 119, 145501 on October 6, 2017}

\begin{abstract}
Which density functional is the "best" for  structure simulations of a 
particular 
material? A concise, first principles, approach to answer this 
question is presented. The random phase approximation (RPA)---
an accurate many body theory--- is used to evaluate various 
density functionals. To demonstrate and verify the method, we apply it to the 
hybrid 
perovskite MAPbI$_3$, a promising new solar cell material. The 
evaluation is done by first creating finite temperature ensembles for small 
supercells using RPA molecular dynamics, and then evaluating the 
variance 
between  the RPA and various approximate density functionals for these 
ensembles. 
We find that, contrary to recent suggestions, van der Waals functionals do not 
improve the description of the material, whereas hybrid functionals and the 
SCAN (strongly constrained appropriately normed) density functional 
yield very good agreement with the RPA. Finally, our study shows that in the 
room temperature tetragonal phase of MAPbI$_3$, the molecules are  
preferentially parallel to the shorter lattice vectors but reorientation on ps 
time scales is still possible.
\end{abstract}

\maketitle

There is no question that Density Functional Theory 
(DFT)\cite{Hohenberg:pr64,Kohn:pr65} has revolutionized our modeling of 
condensed matter and materials
over the last five decades, and it is well understood why this is so: DFT offers
a unique balance between computational efficiency and accuracy--- a balance that 
is simply yet unreached
by any other computational method. Equally important is that forces between the 
atoms are 
readily computable, so that relaxed ground state structures,
vibrational properties, as well as finite temperature properties of any
material are straightforwardly obtainable.  But DFT is not without 
shortcomings. 
For instance, most semi-local functionals completely neglect van der Waals forces, and 
in many open structures such as perovskites the instabilities of the cage are 
not very well described. Quite generally, 
for materials modeling we are often left with the question of 
choosing the right density functional. This requires either some chemical 
intuition,
or better, higher level reference calculations. In this letter, we present a 
fully 
\textit{ab-initio} method based on the Random Phase Approximation 
(RPA) to select the best functional for finite temperature structure 
predictions of a 
certain material (here hybrid perovskites). 

The RPA is an approximate many-body technique for total energies summing
the so called bubble or ring diagrams to infinite order. Since it is 
closely related to
the $GW$ approximation of Lars Hedin~\cite{Hedin:pr65}, it 
should describe band gap related properties very well. By comparison to 
experimental data and diffusion Monte Carlo calculations it has 
been shown that the RPA captures energy differences involving very 
diverse bonding types
from covalent, over ionic, to  van der Waals like very 
accurately~\cite{Marini:prl06,Harl:prl09,Lebegue:prl10,Schimka:natm10,
Olsen:prl11,Bjorkman:prl12,Ren:joms12,Schimka:prb13,Macher:jcp14,Kaltak:prb14}. 
The actual RPA
calculations proceed in two steps. First, a DFT calculation is performed using 
an  approximate density functional, in this case the PBE 
functional~\cite{Perdew:prl96} and all occupied and unoccupied states are 
determined.
Then the RPA correlation energy is evaluated as
\cite{Nozieres:pr58,Langreth:prb77,Miyake:prb02,Fuchs:jcp05,
Furche:jcp08,Harl:prl09}:
 \begin{equation}
 \nonumber
  E_{\rm RPA} = \frac{1}{2 \pi} \int_0^\infty\Tr[ 
\ln(1-\chi(\im\omega)\operatorname{v})+\chi(\im \omega)\operatorname{v}]\,{\rm 
d} \omega, \label{equ:RPA_energy}
 \end{equation}
where $\chi(\im\omega)$  is the independent particle  polarisability calculated 
using DFT orbitals, $\operatorname{v}$
is the Coulomb kernel and $\im \omega$ the imaginary frequency. The exact 
exchange energy~($E_{\rm x}$) is calculated and added: 
$E_{\rm 
xc}=E_{\rm x}+E_{\rm RPA}$. 

If we trust RPA to be accurate for solids, the best strategy is obviously
i) \textit{to calculate finite temperature ensembles using the RPA}, 
ii)\textit{ calculate total energies of the other functionals for these 
"RPA" distributed configurations}, iii)\textit{ rank the functionals 
based on the energy difference ($\Delta E$) to the RPA}. Calculating the 
ensembles 
using some approximate functional is
also possible, but the explored structural phase space might not overlap with 
the correct higher accuracy RPA, biasing the final result.

\begin{figure*}
    \begin{center}
    \includegraphics[height=60mm,clip=true]{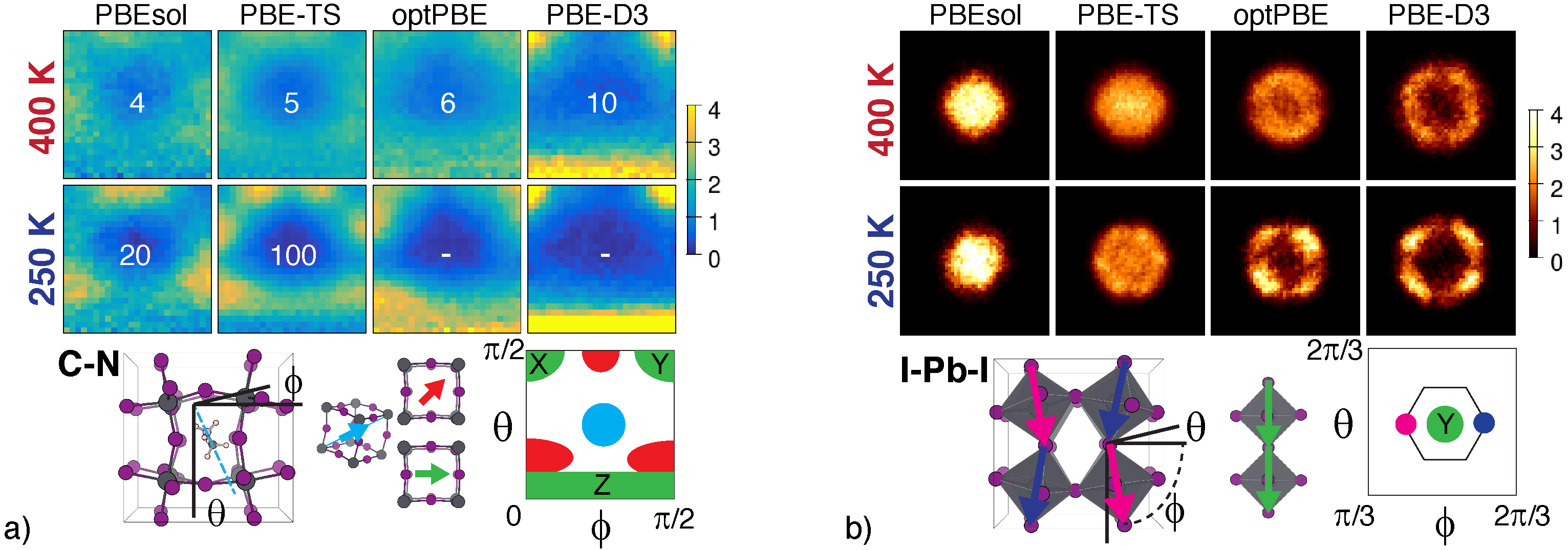}
    \end{center}
   \caption{Polar distribution of the orientation 
$\{\phi,\theta\}$ of 
the a) molecules and b) PbI octahedra (represented by C-N and I-I 
connecting vectors, respectively) in a $2\times2\times2$ supercell as 
function of 
temperature and the XC-functional. For the molecules, cubic symmetry is applied 
to 
down-fold the full polar distribution into a single octant. The molecular 
relaxation times are shown (in ps) by the 
white numbers. The legends below illustrate the orientations  a) of the MA 
molecules in the cage (principle axes -- green, face-diagonal -- red, and 
room-diagonal -- blue) and b) of the I-I connecting vectors (note the different 
values for the range of the angles)
}
\label{fig:orient}
\end{figure*}

This strategy is in principle straightforward, but cumbersome. 
With DFT, we can easily perform finite temperature simulations, yielding 
appropriately
weighted microstates. In the absence of forces, this is rather difficult to do 
for high level methods,
although energies alone suffice to perform  Monte-Carlo (MC) or hybrid 
Monte-Carlo simulations \cite{Nakayama:chl09}.
To achieve high sampling efficiency, the MC moves need to be chosen 
judiciously, 
which is not a simple matter
for systems where the dynamics is not well understood. Finite temperature 
Molecular
Dynamics (MD) are conceptionally simpler, and  this strategy has become
possible with our recent implementation of forces within the RPA
for solid state systems \cite{Ramberger:prl17}. It is also the main strategy 
we adopt here to generate ensembles. How to obtain the RPA forces is elaborated 
in 
detail 
in 
Ref. 
\onlinecite{Ramberger:prl17}. In short, the key point is the insight that the 
first 
derivative of this
functional with respect to the Green's function is just the 
self-energy~($\Sigma$) in the 
$GW$ approximation \cite{Dahlen:pra06}. The relation between the RPA 
correlation energy and the 
$GW$ self-energy is  analogous to the relation between the Kohn-Sham energy and 
its
potential\cite{fn1}.

\begin{figure*}[t]
    \begin{center}
    \includegraphics[height=38.5mm,clip=true]{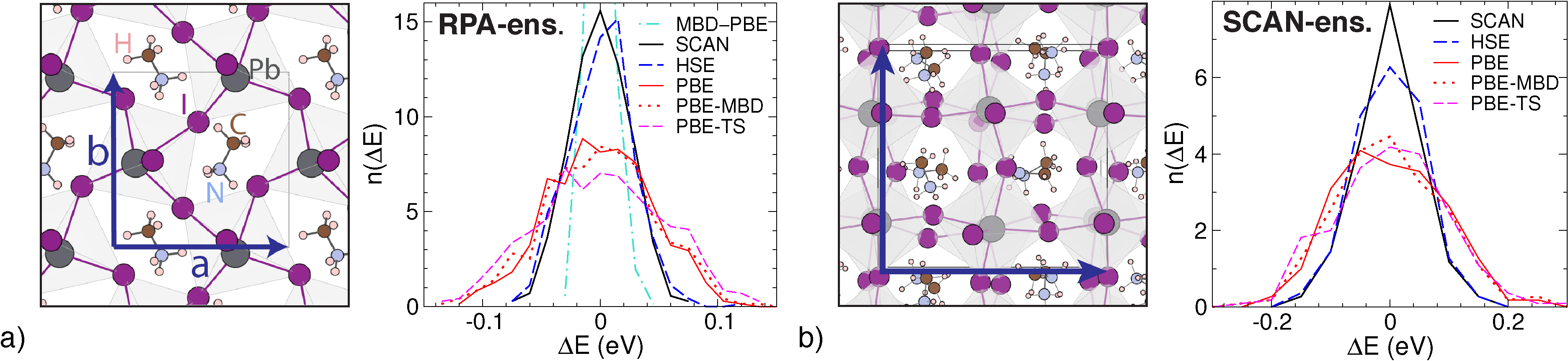} 
    \end{center}
   \caption{a) Snapshot of 
the RPA-MD simulation of MAPbI$_3$, the 
$\sqrt{2}\times{}\sqrt{2}$-cell is indicated by 
the a and b lattice vectors. The histogram presents the energy 
difference between the RPA and selected density functionals for the 
RPA-ensemble. In order to center the histograms the mean 
differences have been subtracted. The $x$-axis shows the energy difference 
(in eV), whereas 
the y-axis indicates the probability of finding a certain energy difference. 
b) Same as  in a) but for the SCAN-ensemble build up with the larger 
$2\times 2\times 2$ cell. Additionally, the difference between the PBE-MBD and 
PBE energies [MBD$-$PBE] is shown in a).
}
\label{fig:RPA}
\end{figure*}

As an example of the predictive power of this method, the experimentally 
difficult to determine atomic structure of hybrid perovskites is studied. This 
is a unique class of 
materials based on an organic donor, methyl-ammonium (MA) CH$_3$NH$_3^+$, 
trapped inside 
a (PbI$_3$)$^-$ perovskite cage~\cite{Green:natpt14}.
These materials have reached
a light to electricity conversion efficiency of 20\%~\cite{NREL16}, approaching 
that of silicon based
solar cells, but they can be produced cheaply using wet chemistry. 
Describing these systems with high precision 
using first principles methods
is very challenging. Instabilities in the perovskite cage  
are often only well captured if band gap related properties and lattice constants
are accurately described. In fact, the MA molecules rotate even at room 
temperature~\cite{Poglitsch:jcp87,Brivio:prb15,*Carignano:jpcc15,*Quarti:pccp15,
*Mattoni:jpcc15}
coupled to concerted movements of PbI octahedra~\cite{Lahnsteiner:prb16}. The 
situation 
is complicated  by 
the molecules being allegedly bonded to the cage
by van der Waals (vdW) interactions~\cite{Egger:jpcl14,Motta:natc15}.

To study the influence of the specific functional on 
the ordering of the molecules and octahedra in MAPbI$_3$ we have performed long
finite 
temperature parallel tempering \textit{ab-initio} MD calculations of 
$2\times2\times2$ pseudo-cubic supercells, 
containing 8 
molecules in PbI$_3$ cages. The calculations were performed 
using 
the {\sc vasp} code\cite{Kresse:prb93,*Kresse:prb96,*Kresse:prb99} with a 
plane-wave basis and the projector augmented wave  (PAW)
method~\cite{Blochl:prb94b} neglecting spin orbit coupling (details in 
Supplemental Material~\cite{SM}). Similarly as in 
Ref.~\onlinecite{Lahnsteiner:prb16}, the 
orientations of all 
molecules inside the cages as well as the orientation of the PbI 
octahedra are extracted in polar coordinates 
$\{\phi,\theta\}$ as a function of time. These data are used to create the 
polar distributions of Figure~\ref{fig:orient}. The 
ordering of the molecules is highly dependent on the functional and even 
amongst vdW functionals there are significant differences. The face-diagonal 
orientations are favored by PBEsol and TS, while optPBE and PBE-D3 favor the 
principle axes. The room-diagonal orientation is avoided by all 
considered functionals. The calculated relaxation 
times of the molecules are shown in Fig.~\ref{fig:orient}~a) as white numbers. 
Most functionals give values in the order of picoseconds (ps) at room 
temperature in reasonable agreement with experiment~\cite{Poglitsch:jcp87}.
The  arrangement of the octahedra can be seen in Fig.~\ref{fig:orient}~b). 
With PBEsol at 250~K the octahedra can tilt and rotate, but without a strong
directional preference. For the other functionals, the octahedra tend to form  
zigzag patterns
(or clockwise anti-clockwise patterns), and avoid orientations exactly along the principle axis. 
At 400~K the orientations are more smeared out. Based 
on these results alone, it is very difficult to determine which 
density functional accurately describes the material.

To judge the quality of each functional, a 
well distributed RPA canonical ensemble of structures is constructed. Because 
of the 
computational cost of the RPA, small $\sqrt{2}\times\sqrt{2}$ cubic cells are 
used. This cell allows
for different orientations of neighboring MA molecules (see Fig. 
\ref{fig:RPA}a). As a starting 
point, we generate independent configurations
using long DFT simulations at our desired temperature of 400~K. 
The internal geometry of the MA molecule is kept rigid 
allowing for a time step of up to 10~fs. For DFT simulations, 
results for rigid MA molecules are practically indistinguishable from 
unconstrained MD for the orientation of the molecules and dynamics of the cage. 
From this trajectory, 16 practically independent configurations are picked
and finite temperature RPA-MD simulations at 400~K are performed. From the last 
1.3~ps of each trajectory, 32 equally spaced configurations 
are picked yielding 512 partly correlated 
configurations. These form the RPA-ensemble. 


\begin{table}[b]
\caption{
Considered functionals, and the square root of
the variance ($\sigma$, in meV) of the energy difference between the RPA
and the functionals in the RPA$|$SCAN$|$PBEsol-ensemble, 
respectively.
}
\label{tab:variance}
\begin{ruledtabular}
\begin{tabular}{lcc|c|c}

{\footnotesize SCAN}& {\footnotesize Strongly Const. Approp. 
Normed}~\cite{Sun:prl15} & 25&53 &67 \\
{\footnotesize HSE-D3}  & \footnotesize 
HSE~\cite{Krukau:jcp06}\footnotesize + Grimme 
D3~\cite{Grimme:jcp10} &27&55&68 \\
{\footnotesize HSE}  & {\footnotesize 
Heyd-Scuseria-Ernzerhof}~\cite{Krukau:jcp06} & 27&57 
&69 \\
{\footnotesize PBE}    & {\footnotesize 
Perdew-Burke-Ernzerhof}~\cite{Perdew:prl96}  & 44&87 
&104 \\
{\footnotesize PBE-MBD}    & {\footnotesize many body 
dispersion}~\cite{Tkatchenko:prl23}& 47&89 
&109 \\
{\footnotesize PBE-D3}& {\footnotesize Grimme D3, vdW 
corrections}~\cite{Grimme:jcp10}& 
51&88 &111\\
{\footnotesize PBE-TS}&{\footnotesize Tkatchenko-Scheffler, vdW 
corrections}~\cite{Tkatchenko:prl09}& 54&90 &109 \\
{\footnotesize optPBE} & {\footnotesize vdW DFT functional based on 
PBE}~\cite{Klimes:prb11}   
& 54&95 &104 \\
{\footnotesize PBEsol} & {\footnotesize revised PBE for 
solids}~\cite{Perdew:prl08} & 55 & 
96&106 
\end{tabular}
\end{ruledtabular}
\end{table}


We evaluate the energy difference between each functional and
the RPA for all configurations in the RPA-ensemble and subtract the mean 
difference, the resulting number is hereafter referred to as 
$\Delta{}\rm E$. The histograms of 
$\Delta{}\rm E$ are shown in Fig. \ref{fig:RPA}~a). The 
curves 
follow 
roughly a Gaussian distribution. In order to rank the functionals, 
we have calculated the square root of the variance of the energy differences 
($\sigma$) for many
functionals and collect the values in Tab.~\ref{tab:variance}. Of all
the considered functionals the HSE, HSE-D3 and SCAN functionals perform 
best, whereas the PBE functional is noticeably worse. To confirm that the 
present results carry over to larger supercells and that the short simulation 
times did not bias the results, 
we performed  MD simulations for a larger $2 \times 2 \times 2$ supercell 
using the best and worst preforming functional (SCAN and 
PBEsol). For both functionals an ensemble of structures was made~\cite{SM}.

As before, the energy differences to the RPA are calculated, the $\sigma$ values tabulated in 
Tab.~\ref{tab:variance} and the 
results for SCAN are plotted in Fig. \ref{fig:RPA}~b). The SCAN  and 
PBEsol-ensemble yield a very similar ranking as the 
RPA-ensemble, although for the PBE derived functionals the order varies
somewhat between the ensembles, with the PBEsol-ensemble clearly deviating
more significantly from the RPA-ensemble. For all ensembles, though, the SCAN functional 
is suggested to be the best choice. This indicates that using ensembles created
with approximate density functionals is a cheap and viable alternative,
if the generation of ensembles using the high level functionals is not possible.

One intriguing result is that the approximate vdW corrected 
functionals do not improve systematically upon PBE.
The many body dispersion (MBD)~\cite{Tkatchenko:prl23} 
correction of Tkatchenko {\em et al.} gives a hint why this is so. 
In MAPbI$_3$, the MBD corrections are small as 
indicated by the histogram
for the energy differences [MBD$-$PBE, dashed dotted line in 
Fig.~\ref{fig:RPA}~a)]. The curve exhibits 
a very narrow peak with $\sigma=10~$meV, whereas $\sigma$ is approximately 30~meV
between PBE and other dispersion corrected schemes for the RPA-ensemble.
This implies that the MBD corrections only add a constant shift
to the energy but vary little between configurations. 
The MBD correction is based on the RPA total energy expression for the 
correlation energy.
However, instead of calculating the polarisability from first principles,
it assumes atomic dipole oscillators at each atomic site. 
Since the PbI$_3$ cage has a small optical gap and screens excitations fairly 
well~\cite{Bokdam:sr16}, dispersion interactions are seemingly strongly washed 
out and show little structure dependence in MBD. The more 
conventional vdW corrections, 
such as TS and  D3, which do not account for any screening of the cage, perform 
worse. For HSE-D3, the damping function in D3 is much stronger than
for PBE reducing the vdW interactions and the structure dependence. 
This explains why HSE-D3 is within the statistical error bars
identical to the parent functional HSE.
Since MBD is by construction more accurate than pairwise 
vdW 
interactions, the virtually perfect agreement 
between PBE and PBE-MBD shows that standard pair-wise 
long range dispersion interactions should not be included for MAPbI$_3$ and, 
presumably, 
most other hybrid  perovskites.

The best means to reduce the error of the semi-local functionals is to 
include exact exchange or use the SCAN functional.
The SCAN functional fulfills all known constraints for DFT functionals 
and depends not only on the density but also on the kinetic 
energy density~\cite{Sun:prl15}. We believe that the main reason for
the improved performance of these functionals is a more balanced description of the cage 
instabilities. The close agreement between HSE and SCAN  
is in line with a recent publication  of Sun {\em et al.} indicating that instabilities 
of perovskites are described equally well by hybrid functionals and 
SCAN~\cite{Sun:natch16}. Remarkably, in the present case, the SCAN functional 
performs even slightly better than HSE. This is most likely related to the SCAN 
functional including short range dispersion effects, which conventional hybrid 
functionals do not account for. We conclude that
SCAN is the most suitable functional to study the atomic structure of hybrid
perovskite materials.

\begin{figure}
    \begin{center}
    \includegraphics[width=80mm,clip=true]{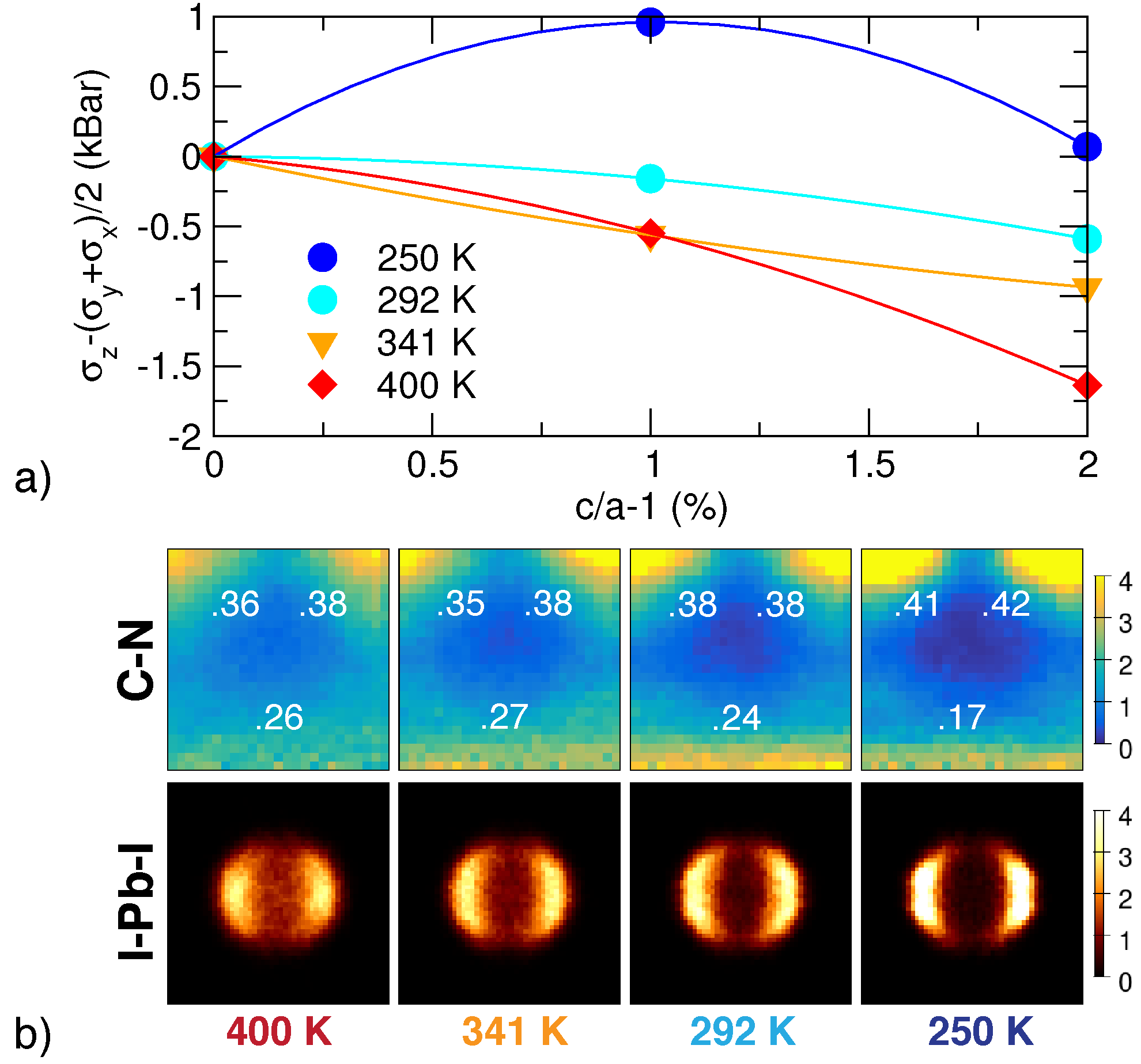} 
    \end{center}
   \caption{a) 
Stress difference in the tetragonal system as function 
of $c/a$-ratio and temperature. b) Polar 
distributions of the MA molecules and PbI octahedra orientations obtained with 
$c/a$=1.01. Only I-I connecting vectors in the $ab$-plane are included. 
Fraction of the molecules aligned along the $a$, $b$ and $c$ axes are 
indicated by white numbers.}
\label{fig:SCAN}
\end{figure}

Last, we use the SCAN functional to study the 
tetragonal distortion ($c/a > 1.0$) that is experimentally observed 
up to and about room temperature\cite{Whitfield:sr16}. To show that this 
distortion is preferred
at low temperatures, we performed parallel tempering MD simulations
for different $c/a$ ratios at a fixed volume ($V=245~$\AA$^3$ per formula unit) 
and calculate the average
stress tensor $\sigma$. The results are shown in \ref{fig:SCAN}~a).
For a stable cubic structure, one would expect that, as one elongates the cell in $c$ 
direction at fixed volume,
a restoring stress opposing the distortion should develop 
($\sigma_z-(\sigma_x+\sigma_y)/2<0$).
However, at the lowest temperature and $c/a=1.01$, the MD suggests
that even further expanding into the $c$ direction is preferable, reaching
a stable point at $c/a=1.02$.  Even at 
T=292~K the system is soft, with a very small restoring
stress that grows roughly quadratic with displacement. 
The conventional linear elastic behavior is only observed at higher 
temperatures.
In agreement with experiment we
therefore predict a tetragonal instability at low temperatures.
Considering the small supercell, one certainly
cannot  expect perfect agreement with experiment for the transition temperature
or $c/a$ ratio. We note that the PBEsol functional, the only other functional 
we tested, does not yield a similar instability.

To study the origin of the distortion, Fig. \ref{fig:SCAN}~b) shows the polar distributions of the MA molecules and the 
PbI octahedra. In the tetragonal cell, the SCAN functional aligns the molecules 
along the principle axes. The fraction of the molecules 
aligned along the $a$, $b$ and $c$ axes is indicated in white. At low temperatures,
the molecules prefer to lie in the ($a, b$) plane, which is shorter than
the $c$ axis ($c/a=1.01$).  As the temperature rises this preference becomes less 
pronounced. This suggests that the molecules act like springs and 
contract the lattice along the direction they are aligned to.
Clearly this hints at a complex interplay between cage instabilities
and molecule-cage interactions.

{\em In summary}, we have presented a general and predictive scheme
to determine the optimal density functional for a specific materials problem.
The strategy relies on the Random Phase Approximation (RPA), which 
is very accurate for condensed matter systems. 
By creating RPA ensembles and evaluating the variance between
the RPA and a selection of DFT functionals, one can choose 
the best functional for large scale structure predictions. 
This avoids the common trial and error approach where different
functionals are explored until agreement with experiment is obtained.
For the  modeling of hybrid perovskites, 
the important finding of the present work is that 
the SCAN functional~\cite{Sun:prl15} works best compared to 
the RPA. Furthermore, many vdW functionals perform disappointingly, which is in 
contrast to
previous suggestions that vdW interactions need to be included in these 
materials~\cite{Egger:jpcl14,Motta:natc15}. 
We have explained this by the strong screening of the cage. 
From a materials point of view, our present study sheds new light on the tetragonal instability of
MAPbI$_3$ (Fig. \ref{fig:SCAN}). We confirm that 
the MA molecules are dynamically reorienting even in 
the tetragonal phase and show that they prefer to be aligned along the two 
shorter axes, $a$ and $b$, thereby contracting the lattice along these 
directions.

{\em Acknowledgement:} Funding by the Austrian Science Fund (FWF): P 30316-N27 
and the SFB 
ViCoM (F41) is gratefully
acknowledged. Computations were performed on the Vienna Scientific Cluster 
VSC3.

\newpage
\appendix
\section{Supplementary Material}

\subsection{Computational Details}

The first principles calculations were performed using 
the {\sc vasp} code\cite{Kresse:prb93,*Kresse:prb96,*Kresse:prb99} with a 
plane-wave basis and the projector augmented wave  (PAW)
method~\cite{Blochl:prb94b}. Setups varied somewhat between the calculations.
For the RPA and SCAN calculations, we used the following {\em accurate} 
potentials
Pb$\_$d$\_$GW, I$\_$GW, C$\_$s$\_$W, N$\_$s$\_$GW, H$\_$GW. The potentials
treat the $5s^25d^{10}6s^26p^2$ (Pb), the ls
$5s^25p^5$ (I),  $2s^22p^2$ (C) and $2s^22p^3$ (N)
as valence orbitals. For most finite temperature MD's (except those for 
constructing the SCAN ensemble), we used {\em softer} potentials 
where the $5d$ electrons of Pb are frozen
(specifically we used the standard Pb, I, C$\_$s, N$\_$s and H potentials). To 
cross-check the potentials,
we also evaluated the variance between the accurate and soft PAW sets
for PBE as well as optPBE and found small values of $\sigma=15$~meV
for the $2\times 2\times 2$ unit cell. For SCAN, the variance
between the two PAW sets doubled, explaining
why we decided to use the accurate set for the SCAN calculations.
The 
plane-wave cut-offs were typically set to 250~eV for MD simulations using the 
soft set
and 300~eV for MD simulations using the accurate set. 

A $\Gamma$-only Brillouin zone sampling for
the $2\times 2 \times 2$ unit cell yields unconverged results
and inaccurate statistics for the molecular 
orientations \cite{Lahnsteiner:prb16}.
To obtain reasonably converged results,
two k-points in the $2\times 2 \times 2$ unit cell [$\Gamma$
and the k-point $(1,1,1)\frac{\pi}{a}$] are required, corresponding
to 16 k-points in the BZ of the primitive cell. 
All RPA calculations and reported variances use this minimal
set of k-points. For the MD's reported in Figure 1, the k-point 
grid
was further doubled (although this did not change the distributions
noticeably).

For the MD simulations, a Langevin thermostat controls the temperature 
and a "Shake and 
Rattle" algorithm is used to constrain all internal degrees of freedom of 
the 
molecules except for the rotation of the NH$_3$ relative to the 
CH$_3$ group over the C$-$N bond. By applying these constraints the 
rapid hydrogen oscillations are removed and this
allows one to 
increase the time step to 10~fs. Parallel Tempering 
\cite{Sugita:cpl99} is 
applied, meaning that four 
trajectories run in parallel at four fixed temperatures between 250 and 400~K. 
Using
a Metropolis algorithm and the kinetic and potential energies of the 
instantaneous structures, 
temperature swaps  between the trajectories are attempted about
every 50 steps. This  increases the sampling efficiency of the phase space, in 
particular, 
at low temperatures. To 
converge the polar distributions of Figure 1, a minimal 
trajectory of 200~ps and up to 750~ps were necessary. The reorientation times 
were obtained from separate fixed temperature MD trajectories with a 
trajectory length of 200~ps or more. The auto-correlation function was 
calculated and fitted 
to a simple exponential as described in Ref. \onlinecite{Lahnsteiner:prb16}. If 
the auto-correlation function did not go to zero, but reaches a plateau above 
zero, a "-" was placed instead of a number. 

All the $2\times2\times2$ and 
$\sqrt{2}\times{}\sqrt{2}$ supercell simulations (Figures 1 and 2) were 
performed for 
pseudo-cubic 
supercells with lattice constants fixed to $a,b=6.312$ and $c=6.316$~\AA{},
which corresponds to the typical experimental volume at 
400~K~\cite{Stoumpos:ic13}.
The simulations representative of the tetragonal distortion ($\sim$300~K) are 
the only exception. Here we slightly reduced the volume to 
245~\AA$^3$ per 
formula unit and used a $2\times2\times2$ supercell (Figure 3). After 
correcting for the Pullay stress, the pressure calculated with the SCAN 
functional 
at 292~K is zero within the statistical error bars. 
The macroscopic stress tensor is obtained by averaging over the elements 
$\sigma_x,\sigma_y,\sigma_z$ calculated for every structure in the MD 
trajectory.

Variances between different exchange correlation functionals and the RPA for 
the 
RPA- and the SCAN-ensemble were always evaluated using the accurate potentials 
and with a cut-off at 
380~eV to reduce errors related to the finite basis sets. This is higher 
than the suggested cut-off of 320 eV for these potentials. The Kohn-Sham 
orbitals 
have been calculated in two steps. First a standard PBE-DFT 
calculation is 
performed 
to obtain the charge density, hereafter an exact digitalization of the 
Hamiltonian is performed including as many orbitals as there are plane-waves in 
the basis. These orbitals are then used to calculate the exact exchange and the 
RPA correlation energies.

There are no D3 parameters by Grimme for the HSE hybrid functional 
in Ref. \onlinecite{Grimme:jcp10}, therefore we used the scaling and damping 
coefficients of the PBE0 functional for the HSE+D3 
calculations. The HSE is a range separated improvement of the PBE0 functional, 
with the same ratio of exact to PBE exchange in the sort range. Furthermore, it
 is exactly equivalent to PBE0 if the range separator ($\omega$) is set to 
zero.

\subsection{Structures in the SCAN and PBEsol ensembles}
For SCAN, 10 independent 16~ps MD simulations were performed at T= 400~K with 
$2\times2\times2$ pseudo-cubic supercells. Furthermore, we included 
supercells at T=273~K starting from tetragonal and orthorhombic prototype 
structures~\cite{Lahnsteiner:prb16} with the same number of atoms as the 
$2\times2\times2$ cells. A total of 200
configurations were selected from these simulations and together form the 
SCAN-ensemble. 
For PBEsol, 64 configurations were picked from the parallel 
tempering simulation at T=400~K shown Figure 1.


\begin{thebibliography}{50}
\bibitem [{fn1()}]{fn1}%
  \BibitemOpen
  \href@noop {} {}\bibinfo {note} {We note that the computational complexity of
  RPA calculations including forces is linear in the number of k-points and
  cubic in the number of plane waves. This is similar to DFT, however, the
  prefactors are typically a factor 10$^2$ (energies) to 10$^3$ (forces) larger
  than for standard DFT calculations. That calculating forces is about 5-10
  times more expensive than calculating total energies is related to the
  computational costs for evaluating the self-energy $\Sigma(\im t)$. The steep
  cost means that one has to restrict simulations to fairly small supercells
  and few k-points.}\BibitemShut {Stop}%
\end{thebibliography}
\end{document}